# Region growing for multi-route cuts

Siddharth Barman*　　Shuchi Chawla†

**Abstract**

We study a number of *multi-route cut* problems: given a graph $G = (V, E)$ and connectivity thresholds $k_{(u,v)}$ on pairs of nodes, the goal is to find a minimum cost set of edges or vertices the removal of which reduces the connectivity between every pair $(u, v)$ to strictly below its given threshold. These problems arise in the context of reliability in communication networks; They are natural generalizations of traditional minimum cut problems where the thresholds are either $1$ (we want to completely separate the pair) or $\infty$ (we don't care about the connectivity for the pair). We provide the first non-trivial approximations to a number of variants of the problem including for both node-disjoint and edge-disjoint connectivity thresholds. A main contribution of our work is an extension of the region growing technique for approximating minimum multicuts to the multi-route setting. When the connectivity thresholds are either $2$ or $\infty$ (the "2-route cut" case), we obtain polylogarithmic approximations while satisfying the thresholds exactly. For arbitrary connectivity thresholds this approach leads to bicriteria approximations where we approximately satisfy the thresholds and approximately minimize the cost. We present a number of different algorithms achieving different cost-connectivity tradeoffs.

*Computer Sciences Dept., University of Wisconsin - Madison, sid@cs.wisc.edu. Supported in part by NSF award CCF-0643763.

†Computer Sciences Dept., University of Wisconsin - Madison, shuchi@cs.wisc.edu. Supported in part by NSF awards CCF-0643763 and CCF-0830494, and a Sloan Foundation Fellowship.

# 1 Introduction

Finding small cuts in graphs is one of the most fundamental combinatorial optimization problems and there is a large literature on exact and approximate algorithms for various versions of this problem. Cut problems have numerous applications; One of the foremost among these is finding bottlenecks in communication networks. For example, the celebrated max-flow min-cut theorem states that the size of the minimum $s$-$t$ cut in a network is equal to the maximum flow that can be routed between $s$ and $t$. Similar (but weaker) duality theorems hold for more general communication patterns, for example, relating the maximum multicommodity flow to the minimum multicut.

From the point of view of reliability of service in the face of edge or node failures, a natural extension to finding the maximum flow in a network is to find a large flow that is spread out across multiple disjoint paths. Such a flow is called a *multi-route flow*. Multi-route flows can be related back to cuts via Menger's theorem [11]: a pair of terminals in a network admits a $k$-route flow (i.e. is $k$-edge-connected) if and only if the minimum cut between the terminals contains at least $k$ edges. This suggests the following natural question: what is the minimum cost set of edges or vertices the removal of which reduces the connectivity of terminal pairs in the network to below a certain threshold? This is the *minimum multi-route cut problem*.

In this paper we provide approximation algorithms for multi-route cut problems. Like traditional cut problems, multi-route cut problems come in multiple flavors depending on whether we are allowed to remove edges or vertices, the desired connectivity ($s$-$t$ cut, multiway cut, multicut, etc.), or whether the connectivity is in terms of edge-disjoint or node-disjoint paths. We provide constant and logarithmic approximations to several of these variants.

It is easy to see that multi-route cut problems are at least as hard as their 1-route counterparts, but they can sometimes be much harder. For example, as noted in [7], a reduction from (1-route) multiway cut shows that single-source multi-sink 2-route cut is APX-hard, whereas the corresponding 1-route version is equivalent to minimum $s$-$t$ cut and is poly-time solvable. Likewise, we show in Section 6 that the following "red-blue" version of $s$-$t$ $k$-route cut is NP-hard for large $k$.[1] In the red-blue $s$-$t$ cut problem, the edge set is divided into red edges and blue edges; The red edges are associated with certain connectivities and the blue edges with certain costs; The goal is to find an $s$-$t$ with total connectivity below a certain threshold and total cost minimized. This version is equivalent to $k$-route $s$-$t$ cut when all the edge connectivities are polynomially bounded.

Multi-route flows were introduced by Kishimoto [9], and have found a number of applications in communication networks [2, 3, 5]. In a series of papers Kishimoto and others [9, 10, 1] developed efficient algorithms for finding multi-route flows, as well as explored approximate max-flow min-cut theorems in this setting. For example, Bagchi et al. [4] showed a strong duality theorem for multi-route flows and cuts in the single-source single-sink case under a non-standard definition of the cost of a cut. More recently, Bruhn et al. [6] considered the single-source uniform costs version of the problem, that is where each edge has a cost of 1. They showed that the gap between a maximum $k$-route flow and a traditional (1-route) maximum flow is at most a factor of $2(1 - 1/k)$. This in turn implies a simple $2(k - 1)$ approximation for the single-source $k$-route cut problem. Bruhn et al. left open the question of designing sub-polynomial approximation algorithms for multi-route cut problems. Note that unlike for 1-route cut problems, in the multi-route case, the uniform cost assumption is not without loss of generality. In particular, replacing an edge of cost $c$ with $c$ parallel edges of cost 1 each can potentially change connectivity between terminal pairs. Therefore Bruhn et al.'s approximation does not extend to a general single-source multi-route cut problem.

The first non-trivial approximations for general multi-route cut problems were developed by Chekuri and Khanna [7]. Chekuri et al. gave LP-rounding based polylogarithmic approximations for the special case of 2-route cuts. In addition to improving upon their approximation factors we solve the two main open problems mentioned in their work—obtaining a polylogarithmic approximation for the 2-route node-disjoint multicut problem, as well as the first non-trivial approximations for $k$-route cuts with $k \geq 3$. Moreover, while Chekuri and Khanna's algorithms are based on a specialized rounding scheme, a main contribution of our work is to develop a general approach based on region growing to solve multi-route cut problems.

[1] The problem is polynomial time solvable for constant $k$.

### Our results and techniques

We consider a natural LP relaxation for multi-route cut problems and extend the "region growing" technique of Garg, Vazirani, and Yannakakis [8] (see also [12]) to this case, providing improved approximations for several versions of the 2-route cut problem and the first non-trivial approximations for $k$-route cut problems.

In a traditional multicut problem the region growing technique guarantees the existence of a cut around every terminal of cost no more than a logarithmic factor larger than the total contribution to the LP objective of edges strictly inside the cut; a logarithmic bound on the approximation then follows from the disjointness of the cuts constructed. Consider a version of the multi-route cut problem in which every edge has cost either 1 or $\infty$.[2] Then our region growing lemma guarantees the existence of a cut around every terminal that has few infinity-cost edges crossing it, while having cost at most a logarithmic factor larger than the contribution to the LP objective of the 1-cost edges inside the cut.

In a traditional multicut setting, an approximation can be obtained by applying region growing successively at each terminal until all terminal pairs are disconnected; In particular, every region has diameter less than 1 and therefore cannot contain more than one terminal belonging to the same terminal pair. In the multi-route setting there are two problems with this approach. First, our LP relaxation defines $h$ different metrics over the graph, one for each terminal pair. Regions are grown with respect to the metric corresponding to the terminal under consideration. Therefore, we can no longer ensure that no terminal pairs survive within a region, and are forced to recurse within regions. This leads to a further logarithmic loss in the approximation factor. Second, as we remove successive regions from the graph, since we do not remove all the boundary edges (specifically, the infinite cost ones), some paths through these regions survive and it becomes tricky to analyze the final connectivity between terminal pairs.

We are able to overcome all of these difficulties for the case of 2-route cuts, and provide $O(\log^2 h)$ approximations to 2-route multicut and multiway cut, where the previous best known approximations due to Chekuri et al. [7] were $O(\log^2 n \log h)$ and $O(\log n \log h)$ respectively. Here $h$ is the number of terminals, and $n$ is the number of vertices in the graph. Furthermore, while Chekuri et al.'s technique does not extend to the node-disjoint version of 2-route multicut, ours extends easily and naturally giving the same approximation factors.

While our region growing lemma extends to the case of $k$-route problems with arbitrary $k$, overcoming the difficulties outlined above appears to require significantly new techinques. In fact, for general connectivity thresholds $k > 2$, the integrality gap of our LP relaxation can be as large as $k$ (see Section 5.1). We therefore explore bicriteria approximations. Straightforward applications of region growing lead to a $(2, 2h)$ and a $(2h, 2)$ bicriteria approximation, where the first factor refers to the approximation in thresholds, and the second to the approximation in cost. By avoiding overlap between successive cuts more carefully, we show how to obtain a $(6, O(\sqrt{h} \log h))$ approximation. These are the first non-trivial approximations in the $k$-route cut case, for $k \geq 3$. We also consider some special cases of the problem. When $h$ is constant or when all the edges have equal cost, we can obtain a $(4, 4)$ and a $(2, 4)$ approximation respectively. The last result holds even when different terminals have different connectivity thresholds.

While the main focus of this paper is on edge-weighted multi-edge-disjoint-route cuts, all of our al-gorithms and analyses extend with little effort to the node weighted and node-disjoint versions as well. We detail the changes required for the node-weighted node-disjoint version in Section 3.2; The other two combinations are identical.

[2] This version in fact captures arbitrary cost multi-route cut problems without loss of generality.

We summarize our main results in Table 1 below. See Section 2 for precise definitions of the various instances of multi-route cut.

| Problem | Previous best result | Our result |
|---|---|---|
| SS-2-EDRC, SS-2-NDRC | $O(\log n)$ [7] | $O(\log h)$ |
| MW-2-EDRC, MW-2-NDRC | $O(\log n \log h)$ [7] | $O(\log^2 h)$ |
| MC-2-EDRC | $O(\log^2 n \log h)$ [7] | $O(\log^2 h)$ |
| MC-2-NDRC | – | $O(\log^2 h)$ |
| SS-$k$-EDRC | – | $(6, O(\sqrt{h} \ln h))$ |
| SS-$k$-EDRC-Uniform | – | $(2, 4)$ |
| SS-$k$-EDRC (constant $h$) | – | $(4, 4)$ |

Table 1: A summary of our main results. See Section 2 for definitions of the problems. $h$ is the number of terminals, and $n$ is the number of nodes in the graph.

## 2 Problem set-up

Given a graph $G = (V, E)$, a pair of nodes $u, v \in V$ are called $k$-edge-connected if there are $k$ edge-disjoint paths between $u$ and $v$ in $G$, and are called $k$-node-connected if there are $k$ node-disjoint paths between $u$ and $v$ in $G$. In multi-route cut problems our goal is to remove a small number (or more generally a low cost set) of edges or nodes from a given graph so as to reduce the connectivity of given pairs of nodes to below certain thresholds.

Like traditional cut problems multi-route cut problems come in different flavors. We begin by formally defining the most general versions we consider. The input to the multicut version of the edge-disjoint-route-cut problem (MC-EDRC) is a graph $G$ with costs $c_e$ on edges, $h$ pairs of vertices called terminals, $\{(s_1, t_1), (s_2, t_2), \cdots, (s_h, t_h)\}$, and connectivity thresholds, $k_i$ for pair $(s_i, t_i)$. The goal is to produce a minimum cost set of edges $E' \subseteq E$, such that for each $i$, $s_i$ and $t_i$ are at most $(k_i - 1)$-edge-connected in the graph $(V, E \setminus E')$. Note that in the traditional multicut problem $k_i = 1$ for all $i$. In the node-disjoint-route multicut (MC-NDRC) problem the goal is to produce a set of edges $E' \subseteq E$, such that for each $i$, $s_i$ and $t_i$ are at most $(k_i - 1)$-node-connected in the graph $(V, E \setminus E')$. Note that although we will mostly talk about edge weighted versions of the problem, our techniques and analyses extend to the node weighted versions as well.

We further study the following special cases:

- $k$-EDRC or $k$-NDRC: here all the connectivity thresholds are equal to a common value $k$.
- 2-EDRC or 2-NDRC: a special case of the above with $k = 2$.
- MW-EDRC or MW-NDRC (MultiWay multi-route cut): we are given a set $T = \{t_1, \cdots, t_h\}$ of terminals with a common connectivity threshold $k$ for every pair $(t_i, t_j) \in T \times T$.
- SS-EDRC or SS-NDRC (Single Source multiple sink multi-route cut): we are given a single source $s$ and a set $T = \{t_1, \cdots, t_h\}$ of terminals with connectivity thresholds $k_i$ for the pair $(s, t_i)$.
- SS-EDRC-Uniform: the version of SS-EDRC where every edge has a cost of 1.

### LP relaxation

The following LP is a relaxation of the MC-EDRC. Other edge-disjoint cut problems have similar LP relaxations. In any integral solution to this LP, edges with $x_e = 1$ are cut, and the (at most) $(k_i - 1)$ edges with $y_e^i = 1$ represent an $s_i$-$t_i$ cut of size at most $(k_i - 1)$ in the residual graph. Note that the LP defines $h$ different shortest path metrics on the graph.

$$\tilde{z} = \min \sum_{e \in E} x_e c_e \qquad \textbf{(ED-LP)}$$

$$\begin{aligned}
\text{s.t.} \quad & \sum_{e \in E} y_e^i \le k_i - 1 && \forall i \in [h] \\
& d^i(u, v) = x_e + y_e^i && \forall i \in [h], e = (u, v) \in E \\
& d^i \text{ is a metric} && \forall i \in [h] \\
& d^i(s_i, t_i) \ge 1 && \forall i \in [h]
\end{aligned}$$

We remark that the algorithms developed by Chekuri et al. [7] were based on a similar but weaker LP. The LP relaxation for the node-disjoint version MC-NDRC is similar (see Section 3.2).

### Notation

We now develop some notation useful in our analysis.

- For a given subset of vertices, $S \subseteq V$, $G[S]$ denotes the subgraph induced by $S$.
- $\mathbf{d}^\ell$ denotes the shortest path metric obtained when edge lengths are given by $\ell_e$. We use $\mathbf{d}^i$ as short-hand for the metric $\mathbf{d}^{x+y^i}$.
- $\mathcal{B}^{\mathbf{d}}(u, r) = \{v \mid \mathbf{d}(u, v) \le r\}$ denotes a ball of radius $r$ around $u$ under metric $\mathbf{d}$. We use $\mathcal{B}^i$ as short-hand for $\mathcal{B}^{\mathbf{d}^i}$.
- For a set $S \subset V$, $\delta(S) = \{(u, v) \mid (u, v) \in E, \ |S \cap \{u, v\}| = 1\}$ is the set of boundary edges of $S$.
- For $S \subset V$, $E(S) = \{(u, v) \mid (u, v) \in E, \ |S \cap \{u, v\}| \ge 1\}$ is the set of all edges incident on $S$. We use $E^i(u, r)$ as short-hand for $E(\mathcal{B}^i(u, r))$.
- For a set $S$, the "$k$-cost" of $S$, denoted $\Gamma^k(S)$, is the total cost of all but the $k - 1$ most expensive edges in $\delta(S)$: $\Gamma^k(S) = \min_{F \subseteq \delta(S); |F| \le k-1} \sum_{e \in \delta(S) \setminus F} c_e$.
- Finally, for $\beta > 0$, the "$(\beta, x)$-volume" of a set $S$ measures the total contribution of all the edges incident on the set to the objective function: $\mathcal{V}^{\beta, x}(S) = \beta + \sum_{e \in E(S)} x_e c_e$.

## 3 Region growing for multi-route cuts

Our main tool for constructing approximations to multi-route cut problems is a region growing lemma. The lemma states that given a feasible solution to the program (**ED-LP**) above, we can find a cut with low $2k$ cost.

We begin by presenting the lemma for the edge-disjoint version of the problem. The following subsection shows the modifications necesary to obtain a version of the lemma for the node-disjoint case.

### 3.1 The edge-disjoint case

**Lemma 1** *Let $G = (V, E)$ be a graph with costs $c_e$ on edges and terminals $s$ and $t$, and $x$ and $y$ be vectors of lengths on edges, such that $\mathbf{d}^{x+y}(s,t) \geq 1$ and $\sum_e y_e \leq k-1$. Then there exists a radius $r < 1$ such that for $S = \mathcal{B}^{x+y}(s,r)$, the $2(k-1)$-cost of $S$, $\Gamma^{2(k-1)}(S)$, is no more than $\alpha \mathcal{V}^{\beta,x}(S)$, where $\alpha = 2\ln\left(\mathcal{V}^{\beta,x}(V)/\beta\right)$.*

*Proof:* For ease of exposition, we assume without loss of generality that there exists a small constant $\epsilon$ such that for every edge $e$, $x_e$ and $y_e$ are multiples of $\epsilon$, and $M = 1/\epsilon$ is an integer. We first modify the graph $G$ such that for every edge $e$, $x_e + y_e = \epsilon$ and only one of these values is non-zero. Specifically we break every edge $e$ into $(x_e + y_e)/\epsilon$ parts with costs $c_e$ each; We assign an $x$ value of $\epsilon$ and a $y$ value of 0 to $x_e/\epsilon$ of these parts, and assign a $y$ value of $\epsilon$ and $x$ value of 0 to the remaining parts. It is clear that the new instance still satisfies the constraints in the theorem statement. Also note that while costs $\Gamma^{2(k-1)}$ stay the same as before, volumes decrease, and so it suffices to prove the lemma for this new fragmented version of the graph.

We will consider $M$ balls centered at $s$ and show that one of these satisfies the criteria in the theorem. For $0 \leq i \leq M$, let $B_i = \mathcal{B}^{x+y}(s, i\epsilon)$, $\mathcal{V}_i^x = \mathcal{V}^{\beta,x}(B_i)$, and $\Gamma_i = \Gamma^{2(k-1)}(B_i)$. We also define the "change in volume", $\Delta\mathcal{V}_i^x$, for the $i$th ball as $\Delta\mathcal{V}_i^x = \sum_{e\in\delta(B_i)} x_e c_e$, with $\Delta\mathcal{V}_0^x = \beta$. Note that $\mathcal{V}_i^x \geq \sum_{a=0}^{i} \Delta\mathcal{V}_a^x$. Also the sets $\delta(B_i)$ are disjoint.

We now prove a few statements about how the change in volume relates to the $2(k-1)$-cost of a ball. Let index set $\Omega$ be defined as follows: $\Omega = \{i \mid i \in [M],\ \epsilon\Gamma_i \leq \Delta\mathcal{V}_i^x\}$. The following lemmas show that $\Omega$ is large.

**Lemma 2** *For any $i \in [M]$, $\epsilon\Gamma_i > \Delta\mathcal{V}_i^x$ implies $\sum_{e\in\delta(B_i)} y_e \geq 2(k-1)\epsilon$.*

*Proof:* We prove the contrapositive statement. Say we have $\sum_{e\in\delta(B_i)} y_e < 2(k-1)\epsilon$, thus edges $e$ in $\delta(B_i)$ with $y_e = \epsilon$ is strictly less than $2(k-1)$. Let $E_y$ be the set of such edges. For the rest of the edges in $\delta(B_i)$ the $x$ value in turn is $\epsilon$. Therefore $\Delta\mathcal{V}_i^x = \epsilon\sum_{e\in\delta(B_i)\setminus E_y} c_e$. Since $|E_y| < 2(k-1)$ and therefore $\sum_{e\in\delta(B_i)\setminus E_y} c_e \geq \Gamma_i$, we have $\Delta\mathcal{V}_i^x \geq \epsilon\Gamma_i$. ■

**Lemma 3** $|\Omega| \geq M/2$.

*Proof:* Recall that $\sum_e y_e \leq k-1$. Now consider an index $i \in [M] \setminus \Omega$. Lemma 2 shows that for such an index $\sum_{e\in\delta(B_i)} y_e \geq 2(k-1)\epsilon$. If the number of indices in $[M]\setminus\Omega$ is strictly more than $M/2$, we would have

$$\sum_e y_e \geq \sum_{i\in[M]\setminus\Omega}\ \sum_{e\in\delta(B_i)} y_e > \frac{M}{2}2(k-1)\epsilon = k-1$$

which gives us a contradiction. ■

We also require the following inequality for the cost analysis below.

**Fact 4** *For any sequence of positive numbers: $a_0, a_1, ..a_N$, the following bound holds*

$$\frac{a_1}{a_0+a_1} + \frac{a_2}{a_0+a_1+a_2} + .... + \frac{a_N}{a_0+a_1+....+a_{N-1}+a_N} \leq \ln\left(\frac{a_0+a_1+...+a_{N-1}+a_N}{a_0}\right)$$

Before proving the fact we show how it leads to the theorem. We focus on the set $\Omega$. Let $\sigma(1),\cdots,\sigma(N)$ be the sequence of indices in $\Omega$ with $N = |\Omega|$. For $\sigma(i) \in \Omega$, let $\tilde{\mathcal{V}}_{\sigma(i)}^x = \sum_{a=0}^{i} \Delta\mathcal{V}_{\sigma(a)}^x$. Note that $\tilde{\mathcal{V}}_{\sigma(i)}^x \leq \mathcal{V}_{\sigma(i)}^x$ for all $i$.

Recall that for all $\sigma(i) \in \Omega$ we have $\epsilon\Gamma_{\sigma(i)} \leq \Delta\mathcal{V}^x_{\sigma(i)}$. Suppose there does not exist an index satisfying the required property, that is, $\forall i \in \Omega$ we have $\Gamma_i > \alpha\mathcal{V}^x_i \geq \alpha\tilde{\mathcal{V}}^x_i$. Thus for all $i \in \Omega$ we have $\frac{1}{\epsilon}\Delta\mathcal{V}^x_i \geq \alpha\tilde{\mathcal{V}}^x_i$, or

$$\frac{\Delta\mathcal{V}^x_i}{\tilde{\mathcal{V}}^x_i} > \alpha\epsilon$$

Summing the above inequality for all the indices in $\Omega$ we get

$$\frac{\Delta\mathcal{V}^x_1}{\tilde{\mathcal{V}}^x_1} + \frac{\Delta\mathcal{V}^x_2}{\tilde{\mathcal{V}}^x_2} + ... + \frac{\Delta\mathcal{V}^x_N}{\tilde{\mathcal{V}}^x_N} > \alpha\epsilon N = \frac{\alpha}{2}$$

Here the last statement follows from Lemma 3 by noting that $N \geq M/2$.
On the other hand, setting $a_0 = \tilde{\mathcal{V}}^x_0 = \beta$ and $a_i = \Delta\mathcal{V}^x_i$, we can apply Fact 4 to get the following which gives us a contradiction.

$$\frac{\Delta\mathcal{V}^x_1}{\tilde{\mathcal{V}}^x_1} + \frac{\Delta\mathcal{V}^x_2}{\tilde{\mathcal{V}}^x_2} + ... + \frac{\Delta\mathcal{V}^x_N}{\tilde{\mathcal{V}}^x_N} \leq \ln\left(\frac{\tilde{\mathcal{V}}^x_N}{\tilde{\mathcal{V}}^x_0}\right) \leq \ln\left(\frac{\mathcal{V}^{\beta,x}(V)}{\beta}\right) = \frac{\alpha}{2}$$

It remains to prove Fact 4.
*Proof of Fact 4:* We prove the fact by induction over $N$. For the base case, with positive integers it is true that $\frac{a_1}{a_0+a_1} \leq \ln\left(\frac{a_0+a_1}{a_0}\right)$. This follows from the fact that for positive $x$ we have $\ln(1+x) \geq \frac{x}{1+x}$ (the function values are equal at $x = 0$ and rate of growth of $\ln(1+x)$ is more than the other). We set $x = a_1/a_0$ here.

By induction hypothesis we assume the inequality to fold for $N-1$. Denoting by $S_k$ the sum $\sum_{i=0}^{k} a_k$, we have

$$\frac{a_1}{S_1} + \frac{a_2}{S_2} + ... + \frac{a_{N-1}}{S_{N-1}} \leq \ln\left(\frac{S_{N-1}}{a_0}\right)$$

Now using the fact that $\ln(1+x) \geq \frac{x}{1+x}$ again with $x = a_N/S_{N-1}$ we have $\ln\left(\frac{S_N}{S_{N-1}}\right) = \ln\left(\frac{a_N+S_{N-1}}{S_{N-1}}\right) \geq \frac{a_N}{S_N}$. Adding the two inequalities we get,

$$\ln\left(\frac{S_N}{S_{N-1}}\right) + \ln\left(\frac{S_{N-1}}{a_0}\right) \geq \frac{a_1}{S_1} + \frac{a_2}{S_2} + ... + \frac{a_{N-1}}{S_{N-1}} + \frac{a_N}{S_N}$$

And hence the claim follows,

$$\frac{a_1}{S_1} + \frac{a_2}{S_2} + ... + \frac{a_{N-1}}{S_{N-1}} + \frac{a_N}{S_N} \leq \ln\left(\frac{S_N}{a_0}\right)$$

■

This concludes the proof of the region growing lemma. ■

Note that in the special case of $k = 2$, the above lemma gives a bound on the 2-cost of the region, which is equivalent to leaving out exactly one edge. Therefore we incur no loss in the connectivity threshold in this case.

While the above lemma suffices to construct approximate solutions to the SS-EDRC, for the multicut version we require additional properties from cuts in our algorithms and so need to consider cuts around both $s_i$ and $t_i$ for a terminal pair $(s_i, t_i)$. We therefore develop the following "two-sided" region growing lemma which shows that we can simultaneously find good disjoint cuts for both $s_i$ and $t_i$.

**Lemma 5** *Let $G = (V, E)$ be a graph with costs $c_e$ on edges and terminals $s$ and $t$, and $x$ and $y$ be vectors of lengths on edges, such that $\mathbf{d}^{x+y}(s,t) \geq 1$ and $\sum_e y_e \leq k-1$. Then there exist radii $r_1 < 1$ and $r_2 > r_1$ such that for $S_1 = \mathcal{B}^{x+y}(s, r_1)$, and $S_2 = V \setminus \mathcal{B}^{x+y}(s, r_2)$, we have for $\alpha = 2\ln \mathcal{V}^{\beta,x}(V)/\beta$:*

- $\Gamma^{2(k-1)}(S_1) \leq 2\alpha\mathcal{V}^{\beta,x}(S_1)$*, and,*
- $\Gamma^{2(k-1)}(S_2) \leq 2\alpha\mathcal{V}^{\beta,x}(S_2)$*.*

*Proof:* The proof is nearly identical to that of Lemma 1. Once again we consider balls with radii $\epsilon i$ centered at $s$, and let $\Omega$ denote the index set of balls with few ($< 2(k-1)$) "$y$-edges". As before, the cardinality of this set, $N$, is at least $M/2$. Consider the balls corresponding to the first $N/2$ indices in $\Omega$. A volume argument identical to the one used previously shows that for one of these balls, say $B_i$, we must have $\Gamma_i \leq 2\alpha\mathcal{V}^x_i$, so $r_1 = i\epsilon$.

In order to find $r_2$ we consider the remaining $N/2$ balls in reverse order. That is, set $B'_1 = V \setminus B_N$, $B'_2 = V \setminus B_{N-1}$ and so on. We can again reapply the volume argument to get a set $B'_j$ satisfying the required properties; $r_2$ would then be $(N-j+1)\epsilon$. In particular the $2(k-1)$ cost of the set is no more than $(\beta, x)$-volume inside it (or outside the corresponding ball $B_{N-j+1}$). By construction $r_2 > r_1$, so we are done. ■

Finally, we note that if we are allowed to charge the cost of a cut to the volume of the entire graph and not just of the cut itself, then we can obtain a stronger version of the region growing lemma:

**Lemma 6** *Let $G = (V, E)$ be a graph with costs $c_e$ on edges and terminals $s$ and $t$, and $x$ and $y$ be vectors of lengths on edges, such that $\mathbf{d}^{x+y}(s,t) \geq 1$ and $\sum_e y_e \leq k-1$. Then there exists a radius $r < 1$ such that for $S = \mathcal{B}^{x+y}(t, r)$, the $2(k-1)$-cost of $S$, $\Gamma^{2(k-1)}(S)$, is no more than $2\mathcal{V}^{\beta,x}(\mathcal{B}^{x+y}(t,1))$.*

*Proof:* The proof is similar to that of Lemma 1. Again we consider balls with radii $\epsilon i$ centered at $t$, and let $\Omega$ denote the index set of balls with few ($< 2(k-1)$) "$y$-edges". As before, the cardinality of this set, $N$, is at least $M/2$, with $M = 1/\epsilon$.

We continue to use the same notation. Thus for $0 \leq i \leq M$, we have $B_i = \mathcal{B}^{x+y}(s, i\epsilon)$, $\Gamma_i = \Gamma^{2(k-1)}(B_i)$ and change in volume, $\Delta\mathcal{V}^x_i$, for the $i$th ball as $\Delta\mathcal{V}^x_i = \sum_{e\in\delta(B_i)} x_e c_e$.

Note that by disjointness of successive balls we have the following inequality

$$\sum_{a=1}^{M} \Delta\mathcal{V}^x_a \leq \mathcal{V}^{\beta,x}(\mathcal{B}^{x+y}(t,1))$$

Let $\sigma(1), \cdots, \sigma(N)$ be the sequence of indices in $\Omega$. We have established that for all $\sigma(i) \in \Omega$, $\epsilon\Gamma_{\sigma(i)} \leq \Delta\mathcal{V}^x_{\sigma(i)}$. Combining the last two inequalities we have $\sum_{i=1}^{N} \epsilon\Gamma_{\sigma(i)} \leq \mathcal{V}^{\beta,x}(\mathcal{B}^{x+y}(t,1))$. Since $\epsilon = 1/M$ and $N \geq M/2$ by averaging argument there exists an index $j$ such that $\Gamma_j \leq 2\mathcal{V}^{\beta,x}(\mathcal{B}^{x+y}(t,1))$ and so the claim follows. ■

### 3.2 Region growing for node-weighted node-disjoint-route cuts

We next consider the version of multi-route cut where we are required to produce minimum weight node cuts, and satisfy thresholds on node-disjoint paths. The LP relaxation for the node-disjoint version MC-NDRC is very similar to program (**ED-LP**). Here $\mathcal{P}_i$ is the set of all paths between $s_i$ and $t_i$. Although this LP is exponential in size, it has an equivalent polynomial-size formulation as above.

$$\tilde{z} = \min \sum_{v \in V} x_v c_v \qquad \textbf{(ND-LP)}$$

$$\text{s.t.} \qquad \sum_{v \in V} y_v^i \leq k_i - 1 \qquad \forall i \in [h]$$

$$\sum_{v \in P} (x_v + y_v^i) \geq 1 \qquad \forall P \in \mathcal{P}_i,\ \forall i \in [h]$$

Region growing works almost in the same way for node-disjoint-route cuts as for edge-disjoint-route cuts. Most importantly, we define volumes and volume increments in terms of the boundary vertices of a set rather than in terms of boundary edges. We sketch below the differences in our definitions and argument to incorporate node-disjointness as well as node costs:

- $\mathbf{d}^{x+y^i}$ is the shortest path metric where the length of a path is the sum of the $x_v$ and $y_v^i$ values of vertices present in it (both end points included). As before we use $\mathbf{d}^i$ as short-hand for the metric $\mathbf{d}^{x+y^i}$.
- As before $\mathcal{B}^{\mathbf{d}}(u, r)$ denotes a ball of radius $r$ around $u$ under metric $\mathbf{d}$, and $\mathcal{B}^i$ is short-hand for $\mathcal{B}^{\mathbf{d}^i}$.
- For a set $S \subset V$, the set of boundary vertices of $S$, $\Delta(S)$ is defined as $\{v \in S \mid \exists (u, v) \in \delta(S)\}$ where $\delta(S)$ are the boundary edges of $S$.
- For a set $S$, $\Gamma^k(S)$ denotes the total cost of all but the $k-1$ most expensive vertices in $\Delta(S)$: $\Gamma^k(S) = \min_{F \subseteq \Delta(S); |F| \leq k-1} \sum_{v \in \Delta(S) \setminus F} c_v$.
- For $\beta > 0$, we define the "$(\beta, x)$-volume" of a set $S$ to be the total contribution of all the vertices in the set to the objective function: $\mathcal{V}^{\beta,x}(S) = \beta + \sum_{v \in S} x_v c_v$.
- As in the proof of Lemma 1 we pick an $\epsilon > 0$ that divides all the $x$ and $y$ values, and fragment the graph by breaking each vertex $v$ into $n_v = (x_v + y_v)/\epsilon$ vertices $v^{(0)}, \cdots, v^{(n_v)}$, each with a cost of $c_v$, and with edges $(v^{(a)}, v^{(a+1)})$ for all $a \in [n_v]$. We replace an edge $(u, v)$ in the original graph by edge $(u^{(n_u)}, v^{(0)})$ if $\mathbf{d}^{x+y}(s, u) \leq \mathbf{d}^{x+y}(s, v)$ and by $(v^{(n_v)}, u^{(0)})$ otherwise. Again it is easy to see that this transformation preserves the costs $\Gamma^{2(k-1)}$ of balls around $s$, but decreases volumes.
- We define $B_i$ to be the ball $\mathcal{B}^{x+y}(s, i\epsilon)$.
- Finally we set the incremental volumes $\Delta\mathcal{V}_i^x$ to be $\sum_{v \in \Delta(B_i)} x_v c_v$ and $\mathcal{V}_i^x = \mathcal{V}^{\beta,x}(B_i)$. As before we have $\mathcal{V}_i^x = \sum_{k=0}^{i} \Delta\mathcal{V}_k^x$.

We therefore get the following node-disjoint analog of Lemma 5. The other two lemmas have similar analogues.

**Lemma 7** *Let $G = (V, E)$ be a graph with costs $c_v$ on vertices and terminals $s$ and $t$, and $x$ and $y$ be vectors of weights on vertices, such that $\mathbf{d}^{x+y}(s, t) \geq 1$ and $\sum_v y_v \leq k - 1$. Then there exist radii $r_1 < 1$ and $r_2 > r_1$ such that for $S_1 = \mathcal{B}^{x+y}(s, r_1)$, and $S_2 = V \setminus \mathcal{B}^{x+y}(s, r_2)$, we have for $\alpha = 2 \ln \mathcal{V}^{\beta,x}(V)/\beta$:*

- $\Gamma^{2(k-1)}(S_1) \leq 2\alpha\mathcal{V}^{\beta,x}(S_1)$*, and,*
- $\Gamma^{2(k-1)}(S_2) \leq 2\alpha\mathcal{V}^{\beta,x}(S_2)$.

# 4 $2$-route cuts

We now apply the region growing technique to 2-route cut problems. A key difference from how the technique is used to find (1-route) multicuts is that we are now working with $h$ different metrics and grow successive regions under different metrics. Nevertheless, in the single-source multi-sink case we can use region growing in much the same way as it is used to find (1-route) multicuts: we successively find small cuts around terminals, remove them from the graph, and recurse on the remaining graph. Unfortunately this simple approach does not work for the more general multicut version of the problem. In particular, while for a traditional multicut no region contains two terminals belonging to the same pair, in our setting it can. We therefore cannot simply remove subgraphs and ignore them; we must recursively produce cuts within each subgraph. We show how to do this repeated cutting at most $\log h$ times in each subgraph, leading to a final approximation factor of $O(\log^2 h)$.

## 4.1 Single-source multiple-sink $2$-route cuts

Once again we will focus on the edge-disjoint case; our algorithm and analysis for the node-disjoint case is identical. Recall that program (**ED-LP**) provides a fractional solution $(x, y)$ to the problem with cost $\sum_e c_e x_e$, and $\sum_e y_e^i \leq 1$ for all $i \in [h]$. This fractional solution defines $h$ different shortest-path metrics $\mathbf{d}^i$ with $\mathbf{d}^i(e) = x_e + y_e^i$ for all $i \in [h]$ and $e \in E$.

Our algorithm for SS-2-EDRC is given in Figure 1. The algorithm starts with an optimal fractional solution to the program (**ED-LP**). At every step it picks an arbitrary terminal still connected to the source, uses the region growing lemma to find an appropriate cut around the terminal, and removes the entire cut from the graph. It continues until no terminals are left. Then for every cut found, it puts back in the graph the most expensive edge in the cut.

To analyze the algorithm we first note that for all $i \in [h]$ the vectors $(x, y^i)$ together satisfy the conditions in Lemma 1 with $k = 2$, and moreover, $2\ln\left(\mathcal{V}^{\beta,x}(V)/\beta\right) \leq 2\ln((\beta + \tilde{z})/\beta) = 2\ln(h+1) = \alpha$. Therefore, we can always find a radius satisfying the conditions of step (3) in the algorithm and the algorithm terminates. It remains to prove that the set $E'$ generated by the algorithm is a legitimate 2-route cut, and analyze its cost. We do this next.

**Lemma 8** *Given a graph $G = (V, E)$ with terminal set $T$ let $E'$ be the set of edges selected by algorithm* SS-2EDRC *then in the graph $H = (V, E \setminus E')$ the universal source $s$ is at most $1$-edge-connected to any terminal present in $T$.*

*Proof:* We claim that in graph $H = (V, E \setminus E')$, for any $a$, a path from the sink $s$ to a vertex $v$, contained in partition $S_a$, must cross $e_a^{max}$.

The proof is by induction over $a$. For the base case we consider a vertex $v$ in $S_1$. $S_1$ is a cut separating $v$ and $s$, so any path from $v$ to $s$ must intersect $\delta(S_1) = \delta'(S_1)$. But $\delta'(S_1) \setminus E' = \{e_1^{max}\}$, therefore our claim holds.

By the induction hypothesis we assume that the claim is true for all vertices in all partitions $S_1$ to $S_{a-1}$. Now consider a vertex $v$ in $S_a$. Consider any path $P$ in $H = (V, E \setminus E')$ from $v$ to $s$, and let $e'$ be the first edge (starting from $v$) on $P$ that is contained in $\delta(S_a)$. For the sake of contradiction assume that $P$ does not contain $e_a^{max}$, so $e' \neq e_a^{max}$. This implies $e' \notin \delta'(S_a)$ because $\delta'(S_a) \setminus E' = \{e_a^{max}\}$. Therefore, $e' \in \delta(S_a) \backslash \delta'(S_a)$. This means that $e'$ got removed from consideration when some partition $S_j$ was removed with $j < a$. One of the vertices of $e'$ survived to be included in $S_a$ thus $e' \in \delta'(S_j)$. But the only edge of $\delta'(S_j)$ present in $E \setminus E'$ (that is in $H$) is $e_j^{max}$, so $e' = e_j^{max}$ and $P$ must now go from a vertex inside $S_j$ to $s$ without recrossing $e_j^{max}$. This is contradicted by the induction hypothesis.

---

**Input:** Graph $G = (V, E)$ with costs $c_e$, source $s$, terminals $T = \{t_1, \cdots, t_h\}$, fractional solution $(x, y)$ with $\sum_e y_e^i \leq 1$ for all $i \in [h]$ and $\mathbf{d}^{x+y^i}(s, t_i) \geq 1$ for all $i \in [h]$. $\tilde{z} = \sum_e x_e c_e$ and $\alpha = 2\ln(h+1)$.
**Output:** A set of edges $E'$ of cost at most $\alpha\tilde{z}$ such that for all $i \in [h]$ $s$ and $t_i$ are at most 1-edge-connected in $(V, E \setminus E')$.

---

1. Initialize $T' \leftarrow T$ and $V' \leftarrow V$.
2. Pick an arbitrary terminal $t_i$ from $T'$. For the rest of the iteration we consider lengths of edges only under metric $\mathbf{d}^i$ and the ball $\mathcal{B}^i(t_i, r)$ is defined over $G[V']$.
3. Let $\beta = \tilde{z}/h$. Pick a radius $r_i \in [0, 1)$ such that $\Gamma^2(\mathcal{B}^i(t_i, r_i))$ is no more than $\alpha\mathcal{V}^{\beta,x}(\mathcal{B}^i(t_i, r_i))$.
4. Set $S_i \leftarrow \mathcal{B}^i(t_i, r_i)$ and update $V' \leftarrow V' \setminus S_i$.
5. Let $T'$ be the set of terminals that are connected to $s$ in $G[V']$.
6. Repeat steps (2) to (5) until $T' = \emptyset$.
7. Let the partitions generated in the previous steps be $S_1$ through $S_l$. Let $\delta'(S_i)$ the set of edges crossing $S_i$ and present in $G[V \setminus \cup_{j=1}^{i-1} S_j]$. Return the set $E' = \bigcup_{i=1}^{l} (\delta'(S_i) \setminus \{e_i^{max}\})$, where $e_i^{max}$ is the maximum cost edge in $\delta'(S_i)$.

---

Figure 1: Algorithm *SS-2EDRC*—Algorithm for single-source multi-sink 2-EDRC

To prove the lemma first note that the above claim immediately implies that any terminal contained in some partition $S_j$ is at most 1-connected to $s$ in $H$. Finally we consider terminals $t$ in the final subgraph $G[V \setminus \cup_{j=1}^{l} S_j]$ disconnected from $s$. Consider any path from such a terminal to $s$ in $H$, say $\overline{e}$ is the first edge (starting at $t$) on $P$ which has exactly one of its vertices in some partition $S_i$. Since $t$ is disconnected from $s$ in the final subgraph such an edge must exist. Note that $\overline{e} \in \delta'(S_i)$. The only way that $\overline{e}$ is not in $E'$ is that it is $e_i^{max}$ but for the rest of the path to be in $H$, $P$ must connect a vertex contained in $S_i$ to $s$ without crossing $e_i^{max}$ which by our claim is not possible. Thus terminals which are present in the final subgraph $G[V \setminus \cup_{j=1}^{l} S_j]$ disconnected from $s$ remain disconnected from $s$ in $H$. ■

Finally we can analyze the cost of the solution. Note that by construction the $l$ edge sets $E(S_1)$, $E(S_2)$, $\cdots$, $E(S_l)$ are pairwise disjoint. Therefore, $\sum_{i=1}^{l} \mathcal{V}^{\beta,x}(S_i) \leq \beta l + \sum_{e \in E} x_e c_e \leq \beta h + \tilde{z} = 2\tilde{z}$. The cost of the final set $E'$ generated by the algorithm is exactly $\sum_i \Gamma^2(S_i)$, which is at most $\alpha \sum_i \mathcal{V}^{\beta,x}(S_i) \leq 2\alpha\tilde{z}$ by construction. The theorem below now follows from noting that $\tilde{z}$ is no more than the cost of the optimal 2-route cut.

**Theorem 9** *Algorithm* SS-2EDRC *generates a* $2$*-edge-route cut of cost no more than* $4\ln(h+1)$ *times the optimal.*

## 4.2 $2$-route multicuts

We now consider the multicut version of 2-EDRC. As before our algorithm successively uses region growing to construct cuts around terminals. However, instead of recursing only on the remaining graph as in the single-source case, this time we need to recurse on both the components in the graph. We show below that by constructing the cuts appropriately, the depth of recursion is at most $\log h$, and therefore we can find a 2-route cut of cost no more than $O(\log^2 h)$ times the optimal.

**Input:** Graph $G = (V, E)$ with costs $c_e$, a set of source-sink pairs $T = \{(s_i, t_i)\}$ along with metric weights on edges: $x_e$ and $y_e^i$ (one for each source sink pair in $T$). $\tilde{z} = \sum_e x_e c_e$, $\beta = \tilde{z}/h$, and $\alpha = 2\ln(h+1)$. Also given are global variables $p$ and $E'$. (Initially $p = 0$ and $E' = \emptyset$.)
**Output:** A set of edges $E'$ such that for all $(s_i, t_i) \in T$, $s_i$ and $t_i$ are at most 1-edge-connected in $(V, E \setminus E')$.

1. If $T$ is empty, stop.
2. Pick a source-sink pair $(s_j, t_j)$ from $T$.
3. Find radii $r_1 \in [0, 1)$ and $r_2 \in (r_1, 1]$ such that $\Gamma^2(\mathcal{B}^j(s_j, r_1)) \leq 2\alpha\mathcal{V}^{\beta,x}(\mathcal{B}^j(s_j, r_1))$ and $\Gamma^2(\mathcal{B}^j(s_j, r_2)) \leq 2\alpha\mathcal{V}^{\beta,x}(V \setminus \mathcal{B}^j(s_j, r_2))$. Note that $\mathcal{B}^j(s_j, r_1)$ and $V \setminus \mathcal{B}^j(s_j, r_2)$ do not intersect.
4. Increment the global index count: $p \leftarrow p + 1$.
5. If the number of connected source-sink pairs in $G[\mathcal{B}^j(s_j, r_1)]$ is less than the number of connected source-sink pairs in $G[V \setminus \mathcal{B}^j(s_j, r_2)]$ then the $p$th cut, $S_p$, is chosen to be $\mathcal{B}^j(s_j, r_1)$, otherwise it is chosen to be $V \setminus \mathcal{B}^j(s_j, r_2)$.
6. Let $e_p^{max} = \text{argmax}_{e \in \delta'(S_p)} c_e$, where $\delta'(S_p)$ is defined to be the set of boundary edges of $S_p$ present in the graph in the current recursive call.
7. Update the global set of edges, $E' \leftarrow E' \cup (\delta'(S_p) \setminus \{e_p^{max}\})$.
8. Recurse on $G[S_p]$ with terminal set being the source-sink pairs connected in $G[S_p]$ and on $G[V \setminus S_p]$ with terminal set being the of source-sink pairs connected in it.

Figure 2: Algorithm *MC-2EDRC*—Algorithm for 2-EDRC Multicut

The algorithm for 2-EDRC multicut is given in Figure 2.

We first note that the vectors $(x, y^i)$ satisfy the conditions of Lemma 5 for terminals $(s_i, t_i)$ and therefore we can always find radii $r_1$ and $r_2$ satisfying the conditions in Step (3).

Next we show that the cost of the final set $E'$ is not too large. Let $\mathcal{S} = \{S_1, S_2, \cdots, S_l\}$, where $l$ is the total number of cuts formed. We claim that every cut in $\mathcal{S}$ is contained in no more than $\log h$ other cuts in $\mathcal{S}$. This follows by noting that, by construction, for any two sets $S_a \subset S_b$ in $\mathcal{S}$, the number of terminal pairs in $G[S_a]$ is no more than half the number of terminal pairs in $G[S_b]$. We therefore have the following lemma.

**Lemma 10** *For a given edge $e \in E$ there are at most $\log h$ cuts in $\mathcal{S}$ such that $e \in E(S_i)$.*

*Proof:* Note that cuts in $\mathcal{S}$ form a laminar family that is for $S_i, S_j \in \mathcal{S}$ either one is contained in the other or they do not intersect at all. Now consider the following collection of cuts $\mathcal{S}_e = \{S \mid S \in \mathcal{S},\ e \in E(S)\}$ also write $l_e = |\mathcal{S}_e|$. Since all the cuts in $\mathcal{S}_e$ intersect we have the following chain of containments over them: $S_{\pi(l_e)} \subseteq S_{\pi(l_e-1)} \cdots \subseteq S_{\pi(1)}$, where $\pi(i)$ is the cut index of the $i$th cut. By our earlier argument the length of such a containment chain can be no more than $\log h$. ■

Finally, in order to bound the cost, as before we have $\sum_p \Gamma^2(S_p) \leq 2\alpha \sum_p \mathcal{V}^{\beta,x}(S_p) = 2\alpha(\beta h + \sum_p \sum_{e \in E(S_p)} x_e c_e)$. Unlike in the single-source case, the edges sets $E(S_p)$ are not disjoint, however, by Lemma 10 we have $\sum_p \mathcal{V}^{\beta,x}(S_p) \leq \beta h + \log h \sum_{e \in E} x_e c_e \leq (\log h + 1)\tilde{z}$. Therefore, the cost of our cut is bounded by $O(\log^2 h)$ times $\tilde{z}$.

It remains to prove that we obtain the desired connectivity among terminal pairs; For this we establish the following useful lemma. We say that a pair of vertices $u, v$ are *first separated* by a cut $S_i \in \mathcal{S}$, if $|S_i \cap \{u, v\}| = 1$ and for all $j < i$, $|S_j \cap \{u, v\}| \neq 1$.

**Lemma 11** *Given $S_i \in \mathcal{S}$, let $u, v$ be a pair of vertices first separated by $S_i$. Then any $u$-$v$ path $P$ in $H = (V, E \setminus E')$ must contain $e_i^{max}$.*

*Proof:* The proof is by induction over the cut index $i$. For the base case suppose that $u \in S_1$ and $v \notin S_1$. Now any path $P$ from $u$ to $v$ must contain an edge from $\delta(S_1)$, say $e$ is the first such edge (starting from $u$). When $S_1$ is constructed by MC-2EDRC, the subgraph under consideration is $G$ itself, so $\delta'(S_1) = \delta(S_1)$. The only edge of $\delta'(S_1)$ present in $E \setminus E'$ is $e_1^{max}$ so $e$ must be $e_1^{max}$.

Next we prove the claim for $S_i$. Let $u, v$ be a pair of vertices first separated by $S_i$ such that $u \in S_i$ and $v \notin S_i$. Now for contradiction assume that there exists a path $P$ from $u$ to $v$ in $H = (V, E \setminus E')$ such that $e_i^{max} \notin P$. Now $P$ must contain an edge in $\delta(S_i)$; Say $e = (u', v')$ is the first such edge in $P$ (starting from $u$), with $u' \in S_i$ and $v' \notin S_i$. Again it is easy to see that $e \in \delta(S_i) \setminus \delta'(S_i)$. This implies that by the time $S_i$ was constructed $e'$ had been removed from the graph. Thus $e \in \delta'(S_j)$ for some $j < i$. Since $P$ is in $H$ we have $e = e_j^{max}$.

Next we show that $S_j$ first separates $v'$ and $v$ but by (strong) induction hypothesis there is no path from $v'$ to $v$ that does not contain $e_j^{max}$, which implies that $P$ can not proceed from $v'$ to $v$ in $H$. Say we label the vertices in $P$ as follows $P = u \to u_1 \to u_2 \to ...u' \to v' \to ....v$. Here $u$ through $u'$ are in $S_i$ and $v' \notin S_i$. Now consider the point of time at which the algorithm constructed $S_j$. The graph under consideration at that time was $\overline{G} = (V, E \setminus (\cup_{k=1}^{j-1} \delta'(S_k))$. We know that $e = (u', v') \in E \setminus (\cup_{k=1}^{j-1} \delta'(S_k))$ since it is in $\delta'(S_j)$. Also the path $u \to u_1 \to ... \to u'$ is present in $\overline{G}$; This follows from the fact that all cuts in $\mathcal{S}$ constructed before $S_i$ either contain no vertex of $S_i$ or contain all the vertices in $S_i$. Moreover there is a path between $u$ and $v$ until $S_i$ is constructed (it is the lowest index cut separating $u$ and $v$), so there is path from $v'$ to $v$ in $\overline{G}$. In other words no cut before $j$ separates $v'$ and $v$. Moreover until we get to the construction of $S_i$ the path between $u'$ and $v$ is intact. But $e \in \delta'(S_j)$, so $S_j$ separates $u'$ and $v'$. Thus $S_j$ separates $v'$ and $v$. This implies that $S_j$ first separates $v'$ and $v$, and we are done. ∎

**Corollary 12** *All source-sink pairs $(s_i, t_i)$ in $T$ at most 1-connected in $H = (V, E \setminus E')$.*

*Proof:* If we have some cut in $\mathcal{S}$ separating source-sink pair $(s_i, t_i)$, then we consider the lowest index cut separating $s_i$ and $t_i$; Say it is $S_j$, that is, $S_j$ *first separates* $s_i$ and $t_i$. Then by the previous lemma any path from $s_i$ to $t_i$ must pass through $e_j^{max}$. This by Menger's Theorem implies that $s_i$ and $t_i$ are 1-connected.

Note that there might be a source-sink pair $(s_i, t_i)$ that gets disconnected (MC-2EDRC continues till all the source-sink pair get disconnected) but no cut in $\mathcal{S}$ separates them. We show that such a pair remains disconnected in $H$ hence proving the corollary.

For contradiction assume there is a path $P$ from $s_i$ to $t_i$ in $H$. Since $s_i$ and $t_i$ are disconnected at the end of algorithm's execution, $P$ must contain an edge from $\delta'(S_j)$ for some $j \in [l]$. Say $\tilde{e} = (\tilde{u}, \tilde{v})$ is the first edge (starting from $s_i$) on $P$ such that $\tilde{e} \in \delta'(S_j)$. $P$ is in $H$ so for $\tilde{e}$ to be in $E \setminus E'$ we must have $\tilde{e} = e_j^{max}$. Next we show that $S_j$ first separates $\tilde{v}$ and $t_i$ so by the previous lemma, $P$ can not proceed from $\tilde{v}$ to $t_i$ without crossing $e_j^{max}$ again, giving rise to a contradiction.

Say we label the path as follows $P = s_i \to w_1 \to w_2 \to ... \to \tilde{u} \to \tilde{v} \to ..t_i$. Note that all the edges on $P$ before $\tilde{e} = (\tilde{u}, \tilde{v})$ are present in $H$. Thus none of these edges belong to any $\delta'(S_i)$ for $i \in [l]$. Since $\tilde{e} \in \delta'(S_j)$ and all edges on $P$ between $s_i$ and $\tilde{u}$ are present in $H$ we have that $S_j$ separates $s_i$ and $\tilde{v}$. Then since no cut in $\mathcal{S}$ separates $s_i$ and $t_i$, we have that $S_j$ separates $\tilde{v}$ and $t_i$. Moreover we claim that no cut with a smaller index separates $\tilde{v}$ and $t_i$. To see this, suppose that $S_k$ for $k < j$ separates $\tilde{v}$ and $t_i$. Since $S_k$ does not separate $s_i$ and $t_i$ (no cut does) we have that $S_k$ separates $s_i$ and $\tilde{v}$. But this implies that we must have an edge of $\delta(S_k)$ on every path between $s_i$ and $\tilde{v}$; In particular the segment of $P$ connecting $s_i$ to $\tilde{v}$ must contain an edge from $\delta'(S_k)$. But this contradicts the assumption that $\tilde{e}$ was the first edge on $P$ contained in some $\delta'()$. Thus no path in $H$ connects $s_i$ and $t_i$. ∎

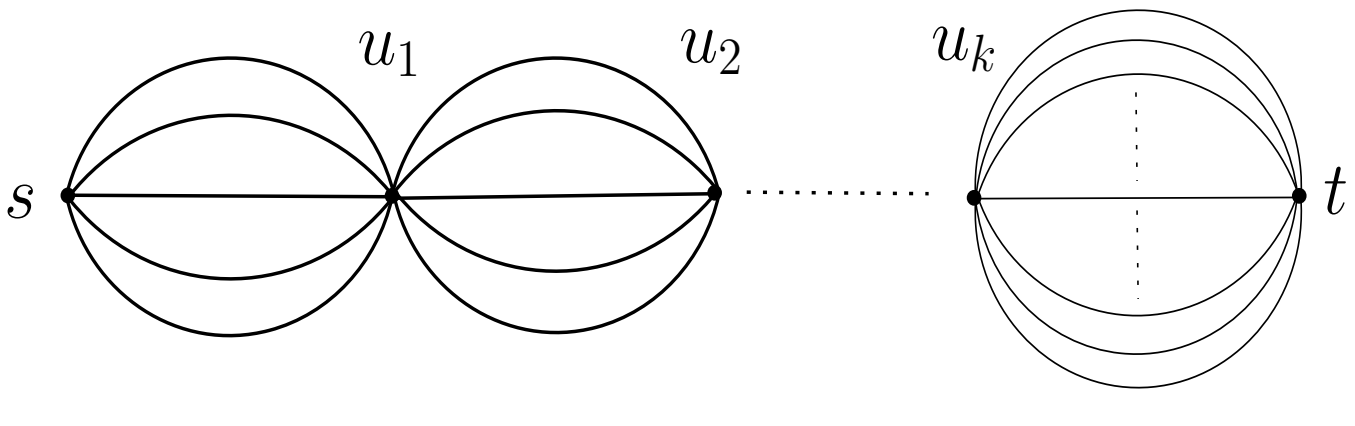

Figure 3: Integrality Gap example

From the cost analysis and Corollary 12 we get the following theorem.

**Theorem 13** *Algorithm* MC-2EDRC *generates a* $2$*-edge-disjoint-route multicut of cost no more that* $O(\log^2 h)$ *times the optimal.*

# 5 $k$-route cuts

We now consider the EDRC and NDRC with larger connectivity thresholds. In Subsection 5.1 it is shown that (**ED-LP**) has a polynomial integrality gap even for the simple case of an $s$-$t$ $k$-EDRC. A similar example can be constructed for (**ND-LP**). Given this large integrality gap, we investigate bicriteria approximations to the EDRC. An $(\alpha, \beta)$ approximation for the $k$-EDRC is a cut of cost at most $\beta$ times the optimal and the removal of which reduces the connectivity between the terminal pair $(s_i, t_i)$ to $\alpha(k_i - 1)$, for every $i$.

## 5.1 Integrality gap for LP (ED-LP)

We present a graph where the optimal integral solution has cost $\Omega(k)$ times the optimal fractional solution. Consider the chain graph in Figure 3 and suppose that we wish to find a $k+1$-route cut separating source $s$ from sink $t$. The graph has $k+1$ parallel edges between $u_i$ and $u_{i+1}$ for all $i \in [k-1]$, each such edge has infinite cost. Also there are $k+1$ infinite cost edges between $s$ and $u_1$. Finally we have $2k$ unit cost edges between $u_k$ and $t$. A feasible fractional solution with cost no more than 2 is obtained as follows: for every edge of infinite cost set $y_e = \frac{1}{k+1}$ and $x_e = 0$, and, for all edges with unit cost, that is edges between $u_k$ and $t$, we set $x_e = \frac{1}{k+1}$ and $y_e = 0$. Note that $\sum_e y_e$ is no more than $k$ and $\sum_e x_e c_e$ is less than 2. Also under the specified edge lengths distance between $s$ and $t_i$ is 1. Hence we have a feasible fractional solution with cost no more than 2. However any integral solution, with finite cost, in order to ensure that the number of edge disjoint paths between $s$ and $t$ is no more than $k$ can only remove $k$ edges between $u_k$ and $t$. Hence an optimal integral cut has cost $k$, giving us an integrality gap of $\Omega(k)$.

## 5.2 The difficulty of applying region growing and some naïve approximations

As mentioned earlier, although the region growing lemma works in the $k \geq 3$ case as well, applying it successively for different terminals leads to the connectivity thresholds being violated by a large factor. Consider, for example, the following algorithm for the single-source $k$-EDRC. We solve **ED-LP**; then for each $i$, we successively apply region growing to the pair $(s, t_i)$ and remove the resulting cut $C_i$ from the graph; our final cut is the collection of all but the $k$ most expensive edges in each $C_i$. The cost of this cut can be bounded by $O(\log h\, \tilde{z})$ using Lemma 1. However, in the final graph, for any terminal $t_j$ with cut $C_j$ there may be several paths to $s$ through cuts $C_i$ for $i < j$ that do not cross $C_j$. Therefore, the best bound we can obtain on the connectivity between $s$ and $t_j$ using this approach is $(k-1)h/2$. In other words, we get an $(O(h), O(\log h))$ approximation.

This approach can be modified slightly to obtain an $(O(h), 2)$ approximation. In particular, we solve the **ED-LP** and combine all the $h$ metrics into a single metric. That is, set $y_e$ to be $\sum_{i=1}^{h} y_e^i$. The metric $d$, defined by setting $d(u,v) = x_e + y_e$ for all edges $e = (u,v)$, separates the source $s$ from all the $h$ terminals — $d(s,t_i) \geq 1$ for all $i \in [h]$. Moreover we have $\sum_{e \in E} y_e \leq kh$. Thus by Lemma 6 we can find a cut $S$ which $2kh-1$ separates $s$ from all the terminals and has cost no more than $2\tilde{z}$. This gives us a $(2h, 2)$ approximation.

One way of avoiding this increase in connectivity is to find successive cuts in the original graph itself, instead of throwing away the previously found cuts. This ensures that connectivity thresholds are maintained to within a factor of 2. However, the cost of the solution can blow up to $O(h\tilde{z})$, implying a $(2, 2h)$ approximation. Specifically, the $(2, 2h)$ approximation is obtained by solving **ED-LP** and applying region growing separately to each pair $(s, t_i)$. Lemma 6 implies that for each terminal $t_i$ we can find a cut $S_i$ which $2(k-1)$ separates the terminal from the source and for which we have $\Gamma^{2(k-1)}(S_i) \leq 2\mathcal{V}^{\beta,x}(\mathcal{B}^{x+y}(t_i, 1))$. Note that $\mathcal{V}^{\beta,x}(\mathcal{B}^{x+y}(t_i, 1))$ is no more than $\tilde{z}$ and so the total cost of the all such cuts is no more than $2h\tilde{z}$. This gives us a $(2, 2h)$ approximation.

### 5.3 Single-source multiple-sink $k$-route cuts

In the remainder of this section, we focus on the single-source case and present a number of different algorithms. The first is a general $(6, O(\sqrt{h}\ln h))$ approximation that relies on a stronger LP (**ED-LP+**) defined below (see Equation (1)). We then consider two special cases — in the first the number of terminals is constant, and in the second all edges in the graph have equal cost. We present a $(4, 4)$ and a $(2, 4)$ approximation for these respectively. These are the first non-trivial approximations for any variant of the $k$-route cut problem with $k \geq 3$.

A key observation that we use for each of these algorithms is that the integral solution to SS-EDRC forms a family of laminar cuts. In particular, let $E'$ be the set of edges removed in an integral solution. By Menger's Theorem we know that for each terminal $t_i$ there exists a set of at most $k_i - 1$ edges whose removal disconnects $t_i$ from $s$ in $(V, E \setminus E')$. Consider any such set of edges, and let $C_i$ be the set of vertices in the connected component containing $t_i$ after these edges have been removed. We call this set a witness for $t_i$. The following lemma shows that for any integral feasible solution we can find a collection of witness sets that are laminar, that is, no two of the sets cross.

**Lemma 14** *For any integral feasible solution to the SS-EDRC there exists a collection of witness sets that is laminar. When all terminals have equal connectivity thresholds, there exists a family of witness sets such that each pair of sets is either identical or disjoint.*

*Proof:* Let $E'$ be an integral solution for the given SS-EDRC. Recall the definition of a witness set. By Menger's Theorem we know that for each terminal $t_i$ there exists a set of at most $k_i - 1$ edges whose removal disconnects $t_i$ from $s$ in $(V, E \setminus E')$. A witness set for $t_i$ is the connected component containing $t_i$ that is formed when we remove any such set of edges from $E \setminus E'$.

Let $H = (V, E \setminus E')$, and note that by definition for any $i \in [h]$ the edge connectivity of $t_i$ and $s$ in $H$ is no more than $k_i - 1$. Of the witness sets for $t_i$ that have the fewest edges crossing them, let $C_i$ be a smallest set in terms of cardinality. We now show that no two smallest witness sets $C_i$ and $C_j$ can cross. Suppose for the sake of contradiction that $C_i$ and $C_j$ cross each other, that is, all three sets $C_i \cap C_j$, $C_i \setminus C_j$ and $C_j \setminus C_i$ are non-empty.

We define the following mutually disjoint sets of edges, here we have $\delta_H(S) = \{(u,v) \in E \setminus E' \mid |\{u,v\} \cap S| = 1\}$

- $O_i = \{(u,v) \in \delta_H(C_i) \mid \{u,v\} \cap C_j = \phi\}$
- $O_j = \{(u,v) \in \delta_H(C_j) \mid \{u,v\} \cap C_i = \phi\}$
- $I_i = \{(u,v) \in \delta_H(C_i) \mid |\{u,v\} \cap C_j| = 2\}$
- $I_j = \{(u,v) \in \delta_H(C_j) \mid |\{u,v\} \cap C_i| = 2\}$

There are three possible cases:

- Suppose that $t_i \in C_i \setminus C_j$ and $t_j \in C_j \setminus C_i$. Then, if $|I_j| < |I_i|$, then $C_i \setminus C_j$ forms a smaller $t_i$-$s$ cut than $C_i$, contradicting the fact that $C_i$ is a witness for $t_i$. Likewise we cannot have $|I_i| < |I_j|$. Therefore $|I_i| = |I_j|$, but then $C_i \setminus C_j$ is a strictly smaller witness set for $t_i$, again contradicting our choice of $C_i$.
- Suppose that $t_i \in C_i \setminus C_j$ and $t_j \in C_j \cap C_i$. This time we must have $|I_i| = |O_j|$ but then $C_i \cap C_j$ forms a strictly smaller witness set for $t_j$.
- Finally, suppose that $t_i, t_j \in C_i \cap C_j$. As before we have $|I_i| = |O_j|$ and $|I_j| = |O_i|$ but then $C_i \cap C_j$ forms a strictly smaller witness set for both $t_i$ and $t_j$.

Therefore the witness sets form a laminar family of cuts.

Note that when all the connectivity thresholds are equal, if there are witness sets $C_i$ and $C_j$ with $C_i \subsetneq C_j$, $C_j$ also forms a witness set for $t_i$. Therefore the lemma holds. ■

#### 5.3.1 An $(6, O(\sqrt{h}\log h))$ bicriteria approximation for single-source cuts

In Figure 4 we present a $\left(6, O(\sqrt{h}\ln h)\right)$ bicriteria approximation algorithm for SS-kEDRC with general edge costs. The algorithm requires an optimal solution an augmented version of **ED-LP**. In particular, we add the following constraint to the LP.

$$\mathbf{d}^i(u,t_i) + \mathbf{d}^j(u,t_j) \geq \mathbf{d}^i(t_i,t_j) \;\; \forall i,j \in [h], u \in V \tag{1}$$

The augmented program is denoted **ED-LP+**. It is easy to see from Lemma 14 that **ED-LP+** is a valid relaxation of the SS-$k$-EDRC. We note that the integrality gap instance of subsection 5.1 applies to this new LP as well. The new constraint is primarily required in Lemma 16 to show that the sets $S$ found in Step 3 (that are constructed under different metrics) are disjoint.

Let us now analyze the algorithm. We first note that we can always find the cuts required for Steps 2a and 3a. For the first, note that if we set $x_e$ and $y_e^i$ to be zero inside $\mathcal{B}^i(t_i, 2/3)$ and scale them up by a factor of 3 outside the ball, then the pair $(s, t_i)$ satisfies the requirements of Lemma 1, and so we can find the desired cut. For the second, if we scale $x_e$ and $y_e^i$ by a factor of 3 inside $\mathcal{B}^i(t_i, 1/3)$ and set them to 0 outside the ball, then again the pair $(s, t_i)$ satisfies the requirements of Lemma 1, and we can find the desired cut.

Next we claim that the connectivity thresholds are satisfied to within a factor of 6. To see this, consider for any terminal $t_i$ the iteration in which $t_i$ is removed from $T$ and let $S$ be the corresponding cut found. Then, $S$ separates $t_i$ from $s$ and we remove all but $6(k-1)$ edges from $\delta(S)$. Therefore our claim follows. Finally, we present a cost analysis. We first show that the algorithm has few iterations.

**Lemma 15** *In each iteration of Steps 2 to 3 the size of $T$ decreases by an additive $\sqrt{T}$.*

**Input:** Graph $G = (V, E)$ with costs $c_e$, source $s$, terminals $T = \{t_1, \cdots, t_h\}$, fractional solution $(x, y)$ that is feasible for **ED-LP+** with connectivity thresholds $k_i = k \; \forall i \in [h]$. $\tilde{z} = \sum_e x_e c_e$ and $\alpha = 2\ln(h+1)$.
**Output:** A set of edges $E'$ of cost at most $O(\alpha\sqrt{h})\tilde{z}$ such that for all $i \in [h]$, $s$ and $t_i$ are at most $6(k-1)$-edge-connected in $(V, E \setminus E')$.

1. Initialize $E' \leftarrow \emptyset$. Let $\beta = \tilde{z}/h$. Set $T' \leftarrow T$.
2. If there is a terminal $t_i \in T'$ such that $|T' \cap \mathcal{B}^i(t_i, 2/3)| \geq \sqrt{|T'|}$, do:
   (a) Pick a radius $r_i \in [2/3, 1)$ with $S = \mathcal{B}^i(t_i, r_i)$ such that $\Gamma^{6(k-1)}(S) \leq 3\alpha\mathcal{V}^{\beta,x}(S)$.
   (b) Let $F(S)$ be the $6(k-1)$ most expensive edges in $\delta(S)$. Set $E' \leftarrow E' \cup (\delta(S) \setminus F(S))$; $T \leftarrow T \setminus S$; $T' \leftarrow T' \setminus S$.
3. Otherwise, while $T' \neq \emptyset$, do:
   (a) Pick a terminal $t_i \in T'$, and a radius $r_i \in [0, 1/3)$ with $S = \mathcal{B}^i(t_i, r_i)$ such that $\Gamma^{6(k-1)}(S) \leq 3\alpha\mathcal{V}^{\beta,x}(S)$.
   (b) Let $F(S)$ be the $6(k-1)$ most expensive edges in $\delta(S)$. Set $E' \leftarrow E' \cup (\delta(S) \setminus F(S))$; $T \leftarrow T \setminus S$; $T' \leftarrow T' \setminus \mathcal{B}^i(t_i, 3/4)$.
4. If $T \neq \emptyset$, set $T' \leftarrow T$ and go to Step 2, otherwise return the cut $E'$.

Figure 4: Algorithm *SS-kEDRC*—Algorithm for single-source multi-sink $k$-EDRC

*Proof:* If Step 2 is executed the lemma follows immediately. Otherwise, note that in each inner loop of Step 3 we remove at most $\sqrt{T'}$ terminals from $T'$. So the loop gets executed at least $\sqrt{T'}$ times. Each time we decrease the size of $T$ by at least 1. Therefore the lemma follows. ■

A simple consequence of this lemma is that the algorithm has at most $O(\sqrt{h})$ iterations. The following lemma bounds the cost of a single iteration and completes the analysis.

**Lemma 16** *In any execution of Step 2 or Step 3 of the algorithm the total cost of the edges included in $E'$ is no more than $6\alpha\tilde{z}$, where $\tilde{z}$ is the value of the* **ED-LP+**.

*Proof:* If Step 2 is executed then the total cost of the edges included is $\Gamma^{6(k-1)}(S)$ which is no more than $3\alpha\mathcal{V}^{\beta,x}(S)$, which in turn is bounded by $3\alpha\tilde{z}$.

Next consider Step 3, and let $S_1, \cdots, S_{h'}$ be the collection of cuts constructed in a single execution of this step. Then we have that the cost of the edges removed in this step is at most $3\alpha\sum_j \mathcal{V}^{\beta,x}(S_j)$. We claim that the sets $S_j$ are disjoint which implies that $\sum_j \mathcal{V}^{\beta,x}(S_j) \leq 2\tilde{z}$, and the total cost for this step is bounded by $6\alpha\tilde{z}$.

To prove the claim, suppose that there are two sets $S_1$ and $S_2$, corresponding to terminals $t_1$ and $t_2$, picked in Step 3 such that $S_1 \cap S_2 \neq \emptyset$. Then for some $u \in S_1 \cap S_2$, $\mathbf{d}^1(u, t_1) \leq 1/3$, and $\mathbf{d}^2(u, t_2) \leq 1/3$, but $\mathbf{d}^1(t_1, t_2) > 2/3$. This directly contradicts constraint (1) in **ED-LP+**. ■

We therefore get the following theorem.

**Theorem 17** *Algorithm SS-kEDRC gives a $\left(6, O(\sqrt{h}\ln h)\right)$ bicriteria approximation for the SS-$k$-EDRC.*

---

**Input:** Graph $G = (V, E)$ with costs $c_e$, source $s$, terminals $T = \{t_1, \cdots, t_h\}$, a partition $\mathcal{P}$ with $l$ sets over terminals along with fractional solution $(x, y)$ satisfying **ED-LP-Part**.
**Output:** A set of edges $E'$ of cost at most $4\tilde{z}$ such that for all $i \in [h]$ $s$ and $t_i$ are at most $4(k-1)$-edge-connected in $(V, E \setminus E')$.

---

1. Double the value of $x_e$ and $y_e^i$ for all edges and for all $i \in [l]$.
2. Repeat for all $1 \leq i \leq l$:
   (a) Construct meta node $v_i$ by merging all terminals in partition set $P_i$.
   (b) Find a cut $S$ separating $v_i$ from $s$ contained in $\mathcal{B}^i(v_i, 1)$ that satisfies $\Gamma^{4(k-1)}(S) \leq 2\mathcal{V}^{\beta,x}(\mathcal{B}^i(v_i, 1))$. Here $d^i()$ is the metric associated with $P_i$.
   (c) Set $F(S)$ to be the set of $4(k-1)$ most expensive edges in $\delta(S)$. Update $E' \leftarrow E' \cup (\delta(S) \setminus F(S))$.

---

Figure 5: Algorithm *SS-kEDRC-const*—Algorithm for single-source multi-sink $k$-EDRC with a constant number of terminals.

### 5.3.2 The constant $h$ case

Recall from Lemma 14 that the witness sets for terminals in the SS-$k$-EDRC are disjoint. When the number of terminals is constant, we can guess the "correct" partition of terminals into groups with identical witness sets. Incorporating this information into the linear program, and finding a good $s$-$t$ $k$-route cut for every group gives us a $(4, 4)$ approximation for the SS-$k$-EDRC.

We know from Lemma 14 that for the SS-kEDRC the witness sets of terminals corresponding to any integral solution are laminar. In fact the collection forms a partition, that is there is a collection of $l$ *mutually disjoint* witness sets: $\{C_1, \ldots, C_l\}$, such that each terminal is contained in one of them. The collection imposes a partition on the terminals. Also, in any integral solution, if terminals $t_a$ and $t_b$ are contained in the same witness set, we have $d^a(t_a, t_b) = d^b(t_a, t_b) = 0$; On the other hand if they are in different cuts we have $d^a(t_a, t_b) = d^b(t_a, t_b) = 1$. We denote by $\mathcal{P}$ the induced partition over terminals: $\mathcal{P} = \{P_1, P_2, \ldots, P_l\}$, where $P_j$ is the set of terminals contained in $C_j$.

Next we present a linear program and the associated algorithm (see Figure 5) which if given the partition $\mathcal{P}$ imposed by an integral solution produces a set of edges that $4(k-1)$ separates every terminal from the source and has cost no more than four times the integral solution. When $h$ is constant we can apply the algorithm over all possible partitions and thus achieve a $(4, 4)$ approximation. The linear program essentially determines $l$ metrics, one for each partition in $\mathcal{P}$ and imposes the corresponding separation requirements.

$$\tilde{z} = \min \sum_{e \in E} x_e c_e \qquad \textbf{(ED-LP-Part)}$$

$$\begin{aligned}
\text{s.t} \quad & \sum_{e \in E} y_e^i \leq k-1 && \forall i \in [l] \\
& d^i(u, v) = x_e + y_e^i && \forall i \in [l], e = (u, v) \in E \\
& d^i \text{ is a metric} && \forall i \in [l] \\
& d^i(s, t_a) \geq 1 && \forall i \in [l], \forall t_a \in P_i \\
& d^i(t_a, t_b) \geq 1 && \forall i \in [l], \forall t_a \in P_i, \forall t_b \notin P_i
\end{aligned}$$

**Input:** Graph $G = (V, E)$, set of terminals $T = \{t_1, t_2, ..t_h\}$ with connectivity requirements $k_i$, and a source vertex $s$.
**Output:** A set of edges $E'$ of cost at most 4OPT such that for all $i \in [h]$ $s$ and $t_i$ are at most $2(k_i - 1)$-edge-connected in $(V, E \setminus E')$.

1. Remove all terminals $t_i$ from $T$ that are at most $2(k_i - 1)$ edge connected to $s$ in $G$.
2. Using a standard mincut algorithm find a set of edges $E'$ that disconnects $s$ from every terminal in $T$.

Figure 6: Algorithm *SS-EDRC-Uniform*—Algorithm for single-source multi-sink EDRC with uniform costs

The algorithm is similar to the algorithm for single-source multi-sink 2-EDRC in Section 4.1 but in addition exploits the fact that each of the partitions have a distance of 1 between them in the optimal LP solution. In particular, we employ the improved region growing lemma (Lemma 6) to argue that the total cost is small.

In order to analyze the algorithm, note that by doubling the $x_e$ and $y_e$ values we have ensured that balls centered at different meta nodes $v_i$ constructed in step (3) of the algorithm are disjoint. Since the $x_e$ values are scaled up by two we have the following: $\sum_{i=1}^{l} \mathcal{V}^{\beta,x}(\mathcal{B}^i(v_i, 1)) \leq 2\tilde{z}$. By Lemma 6 we can find a cut $S$ for each meta node $v_i$, separating the terminals in set $P_i$ from $s$. This ensures that $E'$ is a legitimate $4(k-1)$ route cut for all the terminals. Finally we have $\Gamma^{4(k-1)}(S) \leq 2\mathcal{V}^{\beta,x}(\mathcal{B}^i(v_i, 1))$ for all $i \in [l]$. Combining the last two inequalities we get that the total cost of $E'$ is no more than $4\tilde{z}$. Hence the algorithm achieves a $(4, 4)$ bicriteria approximation.

### 5.3.3 The uniform costs case

Next we consider single-source instances with general connectivity requirements (that is, different terminals are associated with different $k_i$), but where every edge has a cost of 1. We give a $(2, 4)$ bicriteria approximation. Our approach is simple: we ignore terminals that are already less than $2(k_i - 1)$ connected to the source; for the rest we use the characterization in Lemma 14 to argue that cost of a minimum (1-route) cut separating each terminal from the source is no more than 4 times that of the minimum multi-route cut. We therefore find and output the minimum 1-route cut. Figure 6 presents the details.

**Theorem 18** *Algorithm* SS-EDRC-Uniform *returns a set of edges $E'$ of cost at most 4OPT such that for all $i \in [h]$, $s$ and $t_i$ are at most $2(k_i - 1)$-edge-connected in $(V, E \setminus E')$.*

*Proof:* Terminals with less than $2(k_i - 1)$ edge connectivity do not influence the correctness, while terminals that are more than $2(k_i - 1)$ connected to $s$ are totally disconnected from $s$. So the claim about connectivity follows.

Now consider an optimal solution $E_{\text{OPT}}$ for the problem, and let $\mathcal{C} = \{C_i\}$ be the collection of witness sets guaranteed by Lemma14. Let $\mathcal{C}'$ be the subcollection of sets $C_i$ such that for all $C_j \in \mathcal{C}$, $C_i \not\subset C_j$. We claim that $\cup_{C_i \in \mathcal{C}'} \delta(C_i)$ is a multicut for $T$ of cost no more than 4OPT. The first part of the claim follows immediately by noting that each terminal in $T$ is contained in some set $C_i \in \mathcal{C}'$ whereas $s \notin \cup_i C_i$.

For the second part of the claim, consider any $C_i \in \mathcal{C}'$; $t_i$ is the terminal associated with this set. Let $E_i^* = E_{\text{OPT}} \cap \delta(C_i)$. Since $t_i$ is at least $2(k_i - 1)$ connected to $s$ in $G$, $|\delta(C_i)| \geq 2(k_i - 1)$. On the other hand, by the feasibility of $E_{\text{OPT}}$, $|\delta(C_i) \setminus E_{\text{OPT}}| \leq k_i - 1 \leq 1/2|\delta(C_i)|$. Therefore, $|E_i^*| \geq 1/2|\delta(C_i)|$. Now, since any two sets $C_i$ and $C_j$ in $\mathcal{C}'$ are disjoint (Lemma 14), any edge $e$ belongs to at most two of the sets $\delta(C_i)$. Therefore, $| \cup E_i^*| \geq 1/2 \sum_i |E_i^*| \geq 1/4 \sum_i |\delta(C_i)|$, or $|E_{\text{OPT}}| \geq 1/4| \cup_{C_i \in \mathcal{C}'} \delta(C_i)|$. ■

### Multiway cut and multicut with uniform costs

Finally we note that the approach taken in Algorithm *SS-EDRC-Uniform* does not work in the case of multiway EDRC or multicut EDRC. In particular, there is a family of instances of the multiway EDRC parameterized by $k$, containing $\sqrt{k}$ terminals, such that each pair of terminals is $2k+1$ connected, and yet the size of the minimum multiway cut is a factor of $\sqrt{k}$ larger than the size of the minimum multiway $k$-EDRC.

The family is described as follows. Let $t_0, \cdots, t_{h-1}$ be the terminals with $h = \sqrt{k}$. There are $k$ parallel edges between $t_i$ and $t_{i+1 \mod h}$ for all $i \in [h]$, and an additional edge for each pair of terminals, for a total of $\Theta(k^{3/2})$ edges. Then any multiway cut must remove all the $\Theta(k^{3/2})$ edges, whereas in order to obtain a multiway $k$-EDRC, it suffices to remove all parallel edges between $t_0$ and $t_{h-1}$, as well as the $O(h^2)$ additional edges, (leaving a "path" from $t_0$ to $t_{h-1}$,) at a cost of $O(k)$.

## 6 NP-Hardness of $k$-route $s$-$t$ cut

In this section we show that a more general version of $k$-route $s$-$t$ cut is NP-hard for large $k$. The *red-blue $k$-route $s$-$t$ cut* problem is defined as follows. We are given a graph $G = (V, E)$ with a source $s$ and sink $t$, and a connectivity threshold $k$. The edge set $E$ is partitioned into red edges, $E_R$ and blue edges, $E_B$. Edges $e$ in $E_R$ have connectivities $k_e$ associated with them and edges in $E_B$ have cost $c_e$ associated with them. The problem is to find an $s$-$t$ cut $C$ such that $\sum_{e\in\delta(C)\cap E_R} k_e \leq k-1$ and the cost $\sum_{e\in E_B\cap\delta(C)} c_e$ is minimized.

We reduce the knapsack problem to the red-blue $k$ route cut problem. In an instance of the knapsack problem we are given a universe of $n$ items along with a size bound $B$. Here item $i$ has value $v_i$ and size $z_i$. The objective is to find a subset $S$ of items such that $\sum_{i\in S} z_i \leq B$ and the value $\sum_{i\in S} v_i$ is maximized. We construct the graph $G$ with $n$ *intermediate vertices* numbered 1 to $n$, one for each item, along with a source $s$ and a sink $t$. We connect the source $s$ to each of the $n$ intermediate vertices with a red edge (that is, $E_R = \{(s,i)\}_{i=1}^n$). Edge $(s,i)$ is associated with connectivity $k_{(s,i)} = z_i$. Similarly we connect the sink $t$ to the intermediate vertices with blue edges with costs $c_{(i,t)} = v_i$.

It follows that finding a $B$-size bounded set $S$ of items that achieves maximum value is equivalent to finding a min-cost $(B+1)$-route cut in the constructed graph. In particular consider a cut $C$ with $t \in C$ and $s \notin C$. Then it is easy to see that the set $C \setminus \{t\}$ is a valid solution to the Knapsack problem. Furthermore the value achieved by this solution is exactly $\sum_i v_i$ minus the cost of the cut $C$. Therefore, minimizing the cost of $C$ is equivalent to maximizing the value of a feasible knapsack solution, and we get the following theorem.

**Theorem 19** *Red-blue $k$-route $s$-$t$ cut is NP-hard.*

We note that red-blue $k$-route $s$-$t$ cut is equivalent to $k$-route $s$-$t$ cut when the connectivities $k_{(u,v)}$ on edges are polynomially bounded. However, the algorithms developed by us apply to this more general version even with arbitrary edge connectivities. In particular, we can formulate a linear program (**ED-LP-RB**) for the red-blue version, that is similar to the one developed for the $k$-route cut problem in Section 2. Primarily here we ensure that only blue edges have non zero $x_e$ values and only red edges have non zero $y_e$ values.

$$\tilde{z} = \min \sum_{e \in E_B} x_e c_e \qquad \textbf{(ED-LP-RB)}$$

$$\begin{aligned}
\text{s.t} \qquad \sum_{e \in E_R} y_e k_e &\leq k - 1 \\
x_e &= 0 && \forall e \in E_R \\
y_e &= 0 && \forall e \in E_B \\
d(u, v) &= x_e + y_e && \forall e = (u, v) \in E \\
d &\text{ is a metric} \\
d(s, t) &\geq 1
\end{aligned}$$

The following lemma is a counter-part to Lemma 6 and shows that we can obtain a $(2, 2)$-bicriteria approximation for the red-blue $k$-route $s$-$t$ cut problem. As before we use $\mathbf{d}^{x+y}$ to denote the shortest-path metric defined by lengths $x_e$ and $y_e$ on edges.

**Lemma 20** *Let $G = (V, E_B \cup E_R)$ be a graph with cost $c_e$ and connectivity $k_e$ values associated with edges in $E_B$ and $E_R$ respectively. Also let $x$ be vectors of lengths on edges in $E_B$ and $y$ be vectors of lengths on edges in $E_R$, such that $\mathbf{d}^{x+y}(s, t) \geq 1$. Then there exists a radius $r < 1$ such that for $S = \mathcal{B}^{x+y}(s, r)$ we have $\sum_{e \in \delta(S) \cap E_R} k_e \leq 2(k - 1)$ and $\sum_{e \in \delta(S) \cap E_B} c_e \leq 2 \sum_{e \in E_B} x_e c_e$.*

*Proof:* We argue along the lines of the proof of Lemma 6. For simplicity, we assume without loss of generality that there exists a small constant $\epsilon$ such that for every edge $e$, $x_e$ or $y_e$, as the case may be, is a multiple of $\epsilon$, and $M = 1/\epsilon$ is an integer. We first modify the graph $G$ such that for every edge $e$, $x_e = \epsilon$ if the edge is blue or $y_e = \epsilon$ if the edge is red. Specifically we break every edge $e \in E_B$ into $x_e/\epsilon$ parts with costs $c_e$ each and every edge $e \in E_R$ into $y_e/\epsilon$ parts with connectivity $k_e$ each. As before we maintain that only blue edges have non-zero $x$ values and only red edges have non-zero $y$ values. It is clear that the new instance still satisfies the constraints in the lemma statement.

We will consider $M$ balls centered at $s$ and show that one of these satisfies the criteria in the lemma. For $0 \leq i \leq M$, let $B_i = \mathcal{B}^{x+y}(s, i\epsilon)$. Note that the edge sets $\delta(B_i)$ are disjoint. Consider index set $\Omega$ of cuts $B_i$ for which connectivity factor is maintained within a factor of two, that is, $\Omega = \{i \mid i \in [M], \sum_{e \in E_R \cap \delta(B_i)} k_e \leq 2(k - 1)\}$.

We claim that $|\Omega| \geq M/2$. To see this, note that for all indices $j$ in $[M] \setminus \Omega$ we have $\sum_{e \in E_R \cap \delta(B_j)} k_e > 2(k - 1)$. Then for such indices, $\sum_{e \in E_R \cap \delta(B_j)} y_e k_e > 2(k - 1)\epsilon$. Noting that the edges sets $\delta(B_i)$ are disjoint, we get the following sequence of inequalities.

$$k - 1 \geq \sum_{e \in E_R} y_e k_e \geq \sum_{j \in [M] \setminus \Omega} \; \sum_{e \in E_R \cap \delta(B_j)} y_e k_e > 2(k - 1)\epsilon |[M] \setminus \Omega|$$

That is, $|[M] \setminus \Omega| < M/2$, and therefore, $|\Omega| \geq M/2$.

Next, denote the cost of edges crossing $B_i$ as $\Gamma_i = \sum_{e \in \delta(B_i) \cap E_B} c_e$. Recall that $x_e = \epsilon$ for all edges $e \in \delta(B_i) \cap E_B$, for all $i \in [M]$. Hence $\sum_{e \in \delta(B_i) \cap E_B} x_e c_e = \epsilon \Gamma_i$. Just considering indices in $\Omega$ we have

$$\sum_{i \in \Omega} \epsilon \Gamma_i \leq \tilde{z}$$

As shown above, the cardinality of $\Omega$ is at least $M/2$. So by an averaging argument there exists an index $i^*$ in $\Omega$ such that $\Gamma_{i^*} \leq 2\tilde{z}$. As $i^* \in \Omega$ we also have $\sum_{e \in E_R \cap \delta(B_{i^*})} k_e \leq 2(k - 1)$. ■

## 7 Open problems

The most important open question related to our work is that of designing sub-polynomial (bicriteria) approximations to $k$-EDRC. We believe that our region growing lemma will prove to be useful in this regard. Another open problem is to prove non-trivial hardness of approximation results. Currently we merely know that $k$-EDRC problems are at least as hard as their 1-EDRC counterparts. However, we suspect that these problems are much harder, especially for $k > 2$.